\documentstyle[epsf]{article}

\newcommand{\bec}{Bose--Einstein condensate}
\newcommand{\res}{resonance}
\def\be{\begin{equation}}
\def\ee{\end{equation}}
\def\bea{\begin{eqnarray}}
\def\eea{\end{eqnarray}}

\begin{document}

\baselineskip=4.2mm
\twocolumn[\hsize\textwidth\columnwidth\hsize\csname @twocolumnfalse\endcsname

\begin{centering}

{\Large \bf Strongly enhanced inelastic collisions in a \bec\ 
near Feshbach \res s\\[2ex]}

{J. Stenger, S. Inouye, 
M.R. Andrews,
 H.-J. Miesner, D.M. Stamper-Kurn, and W. Ketterle }

{\it Department of Physics and Research Laboratory of
Electronics, \\
Massachusetts Institute of Technology, Cambridge, MA 02139}

\vspace*{.5cm}

\begin{minipage}{13cm}
The properties of Bose--Einstein condensed gases can be strongly altered by
tuning the external magnetic field near a Feshbach \res .
Feshbach \res s affect elastic collisions and lead to the
observed modification of the scattering length. However, as we report
here, this is accompanied by a strong increase in the rate of inelastic
collisions. The observed three--body loss rate in a sodium \bec\ 
increased when 
the scattering length was tuned to both larger or smaller values than 
the off--resonant value. This
observation and the maximum measured increase of the loss rate
by several orders of magnitude are not accounted for by theoretical 
treatments. The strong losses impose severe limitations for using 
Feshbach \res s to tune the properties of \bec s.  
A  new Feshbach resonance in sodium at 1195 G was observed.
\end{minipage}

\end{centering}
\vspace*{.5cm}

]

Most of the properties of \bec s in dilute alkali gases are dominated by
two--body collisions, which can be characterized by the s--wave scattering
length $a$. The sign and the absolute value of the scattering length
determine e.g. stability, internal energy, formation rate, size, and collective
excitations of a condensate. Near a Feshbach \res\ the scattering
length varies dispersively \cite{ties:93,moer:95} covering the whole range of
positive and negative values. Thus it should be possible to study strongly
interacting, weakly or non interacting, or collapsing condensates \cite{kaga:97}
all with the same alkali species and experimental setup.

A Feshbach \res\ occurs when the energy of a molecular \mbox{(quasi--)} bound
state is tuned to the energy of two colliding atoms by applying
an external magnetic field.
Such \res s have been observed in a \bec\ of Na$(F\!=\!1, m_F\!=\!+1$) atoms 
at 853 G and 907 G \cite{inou:98,foot1}, and
in two experiments with cold clouds of $^{85}$Rb($F\!=\!2, m_F\!=\!-2$)
atoms at 164 G \cite{cour:98}.
In the sodium experiment, the scattering length $a$ was observed to vary 
dispersively as a function of the magnetic
field $B$, in agreement with the theoretical prediction \cite{moer:95}:

\be
a = a_0 \left( 1+\frac{\Delta}{B_0-B} \right) , \label{a}
\ee

\noindent where $a_0$ is the off--resonant scattering length, and $\Delta$
characterizes the width of the resonance.

In this Letter we report on the observation
of a broad Feshbach \res\ in sodium in the $F\!=\!1, m_F\!=\!-1$ state at
1195 G, and we investigate the strong
inelastic processes accompanying all three sodium \res s, which result in a
rapid loss of atoms while approaching or crossing the resonances with the 
external magnetic field. The losses show an unpredicted dependence on the 
external magnetic field and impose strong constraints on future experiments
exploiting the tunability of the scattering length.

The experimental set--up is very similar to that described in 
\cite{inou:98}. 
Magnetically trapped condensates in the  $F=1, m_F=-1$ state were transferred 
into an optical trap consisting of a focused far off--resonant 
infrared laser beam \cite{stam:98}. 
The atoms could be transferred to the $m_F=+1$ state by an rf--pulse in a
1 G bias field. For the studies of the Feshbach \res s, 
bias fields of up to 1500 G were applied with ramp speeds of up to 1000 G/ms.
Inhomogenities in the bias field exerted a force mainly in the axial direction.
To prevent atoms from escaping, green light from an argon--ion laser 
was focused into two light sheets at the ends of the cigar shaped condensates,
forming repulsive light barriers. 
The resulting trapping frequencies were around 1500 Hz radially and 150 Hz
axially. 
The trap losses were studied by ramping the bias field with different ramp 
speeds to various field values near the \res s. 
The atoms were probed in ballistic expansion after suddenly switching off
the optical trap. The magnetic field was switched off 2 ms later to ensure 
that the high field value of the scattering length was responsible for the 
acceleration of the atoms. 
After 7 -- 25 ms free expansion the atoms were optically pumped into 
the $F=2$ state and observed in absorption using the cycling transition
in a small bias field of 0.5 G.

The number of atoms $N$, the scattering length $a$ and the mean  density 
$\langle n\rangle$ could be obtained from the absorption images. $N$ is 
calculated from the integrated optical density. The mean field energy 
$2\pi\hbar^2a\,\langle n\rangle /m$ is converted into kinetic energy 
$mv^2_{rms}/2$ after switching off the trap. The root--mean--square velocity 
$v^2_{rms}$ can be extracted from the size of the cloud after the time of flight.
As discussed in Ref. \cite{inou:98}, the mean density $\langle n\rangle$
is proportional to $N(Na)^{-3/5}$ for a three--dimensional harmonic oscillator 
potential, and thus the scattering length $a$ scales as

\be
a \sim \frac{v_{rms}^5}{N} . \label{anorm}
\ee

\noindent Normalized to unity far away from the \res , $v_{rms}^5/N$ is 
equal to $a/a_0$. The proportionality factor in eqn.(\ref{anorm}) involves 
the mean trapping frequency which was not accurately measured. However,
we can obtain absolute values by multiplying the normalized scattering length 
with the theoretical value of $a_0$. The off--resonance scattering length
$a_0=2.75\,{\rm nm}$ of sodium at zero field increases to the triplet 
scattering length $a_0=a_T=3.3\,{\rm nm}$ \cite{will:98} at the high
fields of the resonances. Using this value, the mean density is obtained from the 
root--mean--square velocity,

\be
\langle n\rangle = \frac{m^2v^2_{rms}}{4\pi\hbar^2a} .\label{n}
\ee

\noindent The peak density $n_0$ is given by $n_0=(7/4) \langle n\rangle $ 
in a parabolic potential. 

\begin{figure}[htbf]
\epsfxsize=90mm					
 \centerline{\epsfbox{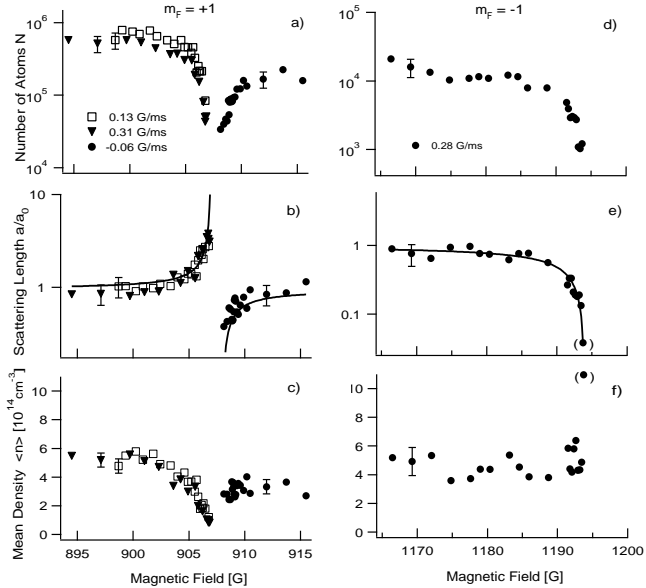}}
\caption{Number of atoms $N$, normalized scattering length
$a/a_0$ and mean density $\langle n\rangle$ versus the magnetic field near the 
907 G and 1195 G Feshbach \res s. The different
symbols for the data near the 907 G \res\ correspond to different ramp 
speeds of the
magnetic field. All data were extracted from time--of--flight images. 
The errors due to background noise of the images, thermal atoms, and loading
fluctuations of the optical trap are indicated
by single error bars in each curve.}
\label{fig1}
\end{figure}

The experimental results for the 907 G and 1195 G 
\res s are shown in Fig.~1.
The number of atoms, the normalized scattering length and the density 
are plotted versus the magnetic field.
The resonances can be identified by enhanced losses. The 907 G
\res\ could be approached from higher field values by crossing it first
with very high ramp speed. This was not possible for the 1195 G \res\ due
to strong losses, and also not for the 
853 G \res\ due to its proximity to the 907 G \res\
and the technical difficulty of suddenly reversing the magnetic field ramp.
For the 907 G \res\ the dispersive change in $a$ can clearly be identified. 
The solid lines correspond to the predicted shapes with width parameters
$\Delta$ = 1 G in Fig.~(b), and $\Delta$ = $-$ 4 G in Fig.~(e). The negative width
parameter for the 1195 G \res\ reflects the decreasing scattering length
when the resonance is approached from the low field side. 
The uncertainties of the positions are mainly due to uncertainties of the
magnetic field calibration. 

The time dependent loss of atoms from the condensate can be parameterized as

\be
\frac{\dot{N}}{N} = - \sum_i K_i\;\langle n^{i-1}\rangle ,\label{Ndoti}
\ee

\noindent where $K_i$ denotes an $i$--body loss
coefficient,  and $\langle n\rangle$ the spatially averaged density.
In general, the density $n$ depends on the number of atoms $N$ in the trap, and
the loss curve is non--exponential. One--body losses, e.g. due to background gas 
collisions or spontaneous light scattering, 
are negligible under our experimental conditions. An
increase of the dipolar relaxation rate (two--body collisions)
near Feshbach \res s has been predicted \cite{ties:93}.
However, for sodium in the lowest energy hyperfine state $F=1, m_F=+1$ 
binary inelastic collisions are not possible. Collisions
involving more than three atoms are not expected to contribute. Thus the
experimental study of the loss processes focuses on the three--body
losses around the $F\!=\!1, m_F\!=\!+1$ \res s, while both two-- and 
three--body losses must be considered for the $F\!=\!1, m_F\!=\!-1$ \res .

Figs.~1 (c) and (f) show a decreasing or nearly constant density when 
the \res s were approached. Thus the enhanced trap losses
can only be explained with increasing coefficients 
for the inelastic processes. The quantity $\dot N/N \langle n^2\rangle$ is 
plotted versus the magnetic field in Fig.~2 for both the 907 G and the 1195 G 
\res s. Assuming that mainly three--body collisions cause
the trap losses, these plots correspond to the coefficient $K_3$. 
For the 907 G (1195 G) \res , the off--resonant value
of $\dot N/N \langle n^2\rangle$ is about 20 (60) times larger than the
value for $K_3$ measured at low fields \cite{stam:98}. Close to the \res s, the
loss coefficient strongly increases both when tuning the scattering length
larger or smaller than the off--resonant value. Since the density is nearly 
constant near the 1195 G \res , the data can also be interpreted by a two--body 
coefficient $K_2=\dot N/N \langle n\rangle$ with an off--resonant value of 
about $30\times 10^{-15}\,{\rm cm^3/s}$, increasing by a factor of more than 
50 near the \res . The contribution of the two processes cannot be distinguished 
by our data. However, dipolar losses are much better understood than 
three--body losses. The off--resonant 
value of $K_2$ is expected to be about $10^{-15}\,{\rm cm^3/s}$ 
\cite{boes:96b} in the magnetic field range around the Feshbach \res s, 
suggesting that the three--body collisions are the dominant loss mechanism
also for the 1195 G \res .

\begin{figure}[htbf]
\epsfxsize=60mm						
 \centerline{\epsfbox{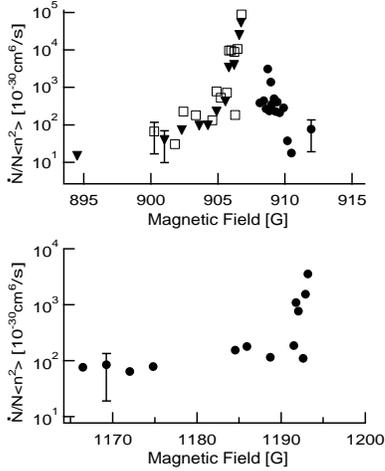}}
\caption{Rate coefficients for inelastic collisions near Feshbach \res s.
$\dot N/N \langle n^2\rangle$ is plotted versus the magnetic field. 
The time derivative $\dot{N}$ was calculated from neighboring points 
without smoothing the data.}
\label{fig2}
\end{figure}

The data analysis following eqns.(\ref{anorm}) and (\ref{n}) is 
based on two major assumptions: the Thomas--Fermi
approximation in which the kinetic energy is small compared to the 
mean--field energy, and the assumption that the atoms maintain the equilibrium 
density distribution during the change of the scattering length.
The Thomas--Fermi approximation is well justified for all data points with 
$a/a_0 > 1$. For $a/a_0 < 1$ close to the \res s, the mean field energy of the 
condensates is only a few times larger than the level spacing in the trap. This 
leads to an overestimate of the density and thus to an underestimate of $K_3$
for the data points with smallest scattering length.
The ramp speeds of the magnetic field were chosen low enough to ensure
adiabaticity, characterized by the condition 
$\dot{a}/a \ll \omega_i$ \cite{kaga:97}, where $\omega_i=2\pi\nu_i$ are the 
trapping frequencies. 
Only for the data point closest to the 1195 G \res\ (in parentheses) 
this condition is not fulfilled.
Indeed, the scattering length $a/a_0$ (Fig. 1b) and the quantity
$\dot N/N \langle n^2\rangle$ (Fig. 2a) are independent of
the ramp speeds, supporting the assumption of adiabaticity. 

The observed increase and the magnetic field dependence of the three-body
collision rate is not accounted for by any theory.  Two theoretical
treatments suggested that the rate coefficient $K_3$ should vary monotonically
with the scattering length following some power law. Fedichev et al.
\cite{fedi:96} derived the universal relation $K_3=3.9\hbar a^4/2m$. This 
prediction was in fairly good agreement with measurements in 
rubidium \cite{burt:97} and sodium \cite{stam:98} in low magnetic fields, even 
though it is not clear that the assumptions are valid in these experiments. 
By considering the
breakup of a dimer by a third atom as the inverse process of recombination,
Moerdijk et al. \cite{moer:96} suggested that the recombination rate is
proportional to $a^2$.
Although the assumptions of those theories might not be fulfilled near
Feshbach resonances, they raised the hope that loss rates should not
increase in the region of the Feshbach \res\ where the scattering
length is small. However, for scattering lengths smaller than $a_0$ a 
substantial increase of the coefficient $K_3$ was observed.
Thus, these measurements show the need for a more accurate theoretical 
treatment of ultracold three--body collisions.

Another way to characterize the losses is a rapid sweep across the \res s.
We determined the fraction of atoms which were lost in sweeping
through the 853 G \res\ and the 907 G \res\ at different 
ramping speeds \cite{footnote}.
For this, the optical trap and the magnetic field were suddenly switched off, 
either before or after crossing the resonance, and the number of atoms was
determined from absorption images as before.

\begin{figure}[htbf]
\epsfxsize=60mm						
 \centerline{\epsfbox{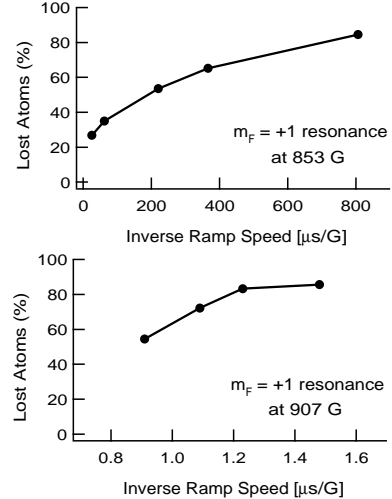}}\vspace{.5cm}
\caption{Fraction of lost atoms after crossing the Feshbach resonances at
853 G and 907 G for different inverse ramp speeds of the magnetic field.}
\label{fig3}
\end{figure}

Fig.~3 shows the fraction of lost atoms as a function of the inverse  ramp 
speed. Using the width of 1 G for the 907 G \res , this implies that 70 \% of all 
atoms are lost in one microsecond.  Across the 853 G \res\ 70 \% of all atoms 
are lost at a ramp speed of $1 \,{\rm G}/400\,{\rm \mu s}$. Assuming a
universal behavior near Feshbach \res s, this implies that the strength of the 
853 G \res\ is four hundred times weaker than of the 907 G \res . Since the 
width $\Delta$ of the Feshbach \res\ is proportional to the 
coupling strength between the two involved molecular states 
\cite{moer:95}, a width of 0.0025 G for 
the 853 G \res\ is estimated. A theoretical prediction for the widths of the
907 G and the 853 G \res s are 1 G and 0.01 G, respectively \cite{abee:98}.

The loss of 70 \% of the atoms in only one microsecond cannot be explained
by the usual picture of inelastic collisions. Due to the very small kinetic energy
of Bose--Einstein--condensed atoms, which is of order
$\frac{\hbar^2}{2m}(\frac{n}{N})^{2/3}$ \cite{park:98}, 
the travel distance of an atom in 1 $\mu {\rm s}$ is less than 1 nm under our 
conditions, much smaller than the mean distance between the atoms, 
$d\approx n^{-1/3}\approx$ 100 nm.  
Loss by two-- or three--body collisions should then be limited to the small 
fraction of atoms which happens to be very close to other atoms.  Possible 
explanations for the observed large losses
are the divergence of the scattering length which allows for 
extremely long--range interactions between atoms, the formation of a 
molecular condensate, as recently proposed \cite{timm:98}, or the impulsive
excitation of solitons \cite{rein:98} when the region of negative
scattering length is crossed.  All suggested explanations require new
concepts in many--body theory.

In conclusion, we have reported on the observation
of a new, $F=1, m_F=-1$ Feshbach resonance at 1195 G in sodium, and we have
investigated the enhanced trap losses accompanying all three resonances
observed so far. The scattering length could be altered in the range of
$0.2 \le a/a_0 \le 5$. The three--body loss coefficient $K_3$ increased
for both larger and smaller scattering length by up to more than three
orders of magnitude.  Those losses show that the density of the sodium
samples must be reduced well below $10^{14}\,{\rm cm^{-3}}$ for further
studies exploiting the tunability of the scattering length near Feshbach 
\res s.

Our sweep experiments revealed that approaching a Feshbach resonance from
the high magnetic field side is strongly affected by trap loss while
crossing the resonance.  The new $m_F\!=\!-1$
\res\ at 1195 G is well suited for studies of \bec s with
zero or negative scattering length, since this region is on the low field
side of the resonance and can be directly approached without crossing any 
resonance.

So far, the physics of gaseous Bose-Einstein condensates has been very well
described including only binary interactions between the atoms \cite{park:98}.
Feshbach \res s might open the possibility to study physics beyond this
approximation which breaks down when the scattering length diverges, and 
also for
very small scattering lengths when one has to consider higher order terms in the
atomic interactions.  Feshbach resonances may also lead to new inelastic
processes.  The observed losses of atoms near Feshbach resonances indicate
molecular and many-body physics which is not yet accounted for by any theory.

This work was supported by the Office of Naval Research, NSF, Joint
Services Electronics Program (ARO), NASA, and the David and Lucile Packard
Foundation. J.S. would like to acknowledge support from the Alexander von
Humboldt--Foundation and D.M.S.-K. from the JSEP Graduate Fellowship Program.


\begin{thebibliography}{999}
\bibitem{foot1}
	Although $F$ is a good quantum number only in small magnetic fields,
	in our case it unambigously defines the states at high fields together 
	with $m_F$.\vspace*{-2.5mm}
\bibitem{footnote}
	The maximum ramp speed was 1000 G/ms. The switch--off time of the
	magnetic field was much shorter than a millisecond. The time between
	the crossing of the 853 G (907 G) resonance and the trap switch off 
	was in the range of 1 -- 20 ms (100 - 400 $\mu$s). The atom density
	had already dropped by a factor of at least ten for the 907 G \res\ and more
	than 100 for the 853 G \res\ before crossing the \res\ again during the
	switch off. Thus additional losses of atoms during the switch--off of
	the magnetic field can be neglected.\vspace*{-2.5mm}
\end{thebibliography}
\end{document}